# Large-Scale Evaluation of Mobility, Technology and Demand Scenarios in the Chicago Region Using POLARIS

**Joshua Auld\*, Jamie Cook, Krishna Murthy Gurumurthy, Nazmul Khan, Charbel Mansour, Aymeric Rousseau, Olcay Sahin, Felipe de Souza, Omer Verbas, and Natalia Zuniga-Garcia[1]**

Transportation and Power Systems (TAPS) Division
Argonne National Laboratory
9700 S. Cass Avenue, Lemont, IL 60439

\* Corresponding Author, Email: jauld@anl.gov

## ABSTRACT

Rapid technological progress and innovation in the areas of vehicle connectivity, automation and electrification, new modes of shared and alternative mobility, and advanced transportation system demand and supply management strategies, have motivated numerous questions and studies regarding the potential impact on key performance and equity metrics. Several of these areas of development may or may not have a synergistic outcome on the overall benefits such as reduction in congestion and travel times. In this study, the use of an end-to-end modeling workflow centered around an activity-based agent-based travel demand forecasting tool called POLARIS is explored to provide insights on the effects of several different technology deployments and operational policies in combination for the Chicago region. The objective of the research was to explore the direct impacts and observe any interactions between the various policy and technology scenarios to help better characterize and evaluate their potential future benefits. We analyze system outcome metrics on mobility, energy and emissions, equity and environmental justice and overall efficiency for a scenario design of experiments that looks at combinations of supply interventions (congestion pricing, transit expansion, tnc policy, off-hours freight policy, connected signal optimization) for different potential demand scenarios defined by e-commerce and on-demand delivery engagement, and market penetration of electric vehicles. We found different combinations of strategies that can reduce overall travel times up to 7% and increase system efficiency up to 53% depending on how various metrics are prioritized. The results demonstrate the importance of considering various interventions jointly.

---

[1] Authors in alphabetic order

## INTRODUCTION

Innovations in vehicle connectivity, automation and electrification and development of new policies for managing congestion, travel demand and network operations, have led to numerous studies that attempt to predict impacts on mobility, energy use, equity, and other key metrics. Many studies to date have explored the impact of individual technologies. While estimating potential impacts independently is already a challenging task given the many unknown aspects of emerging technologies, it is important to understand how these technologies interact with each other as more significant impacts, whether synergistic or antagonistic, may emerge. This is especially important when developing wide-ranging decarbonization strategies to drastically reduce GHG emissions while maintaining or enhancing transportation system improvement, as targeted in the recent blueprint from the U.S. Departments of Energy, Transportation, Housing and Urban Development, and the Environmental Protection Agency [1].

The emergence of technologies like shared mobility, vehicle-to-vehicle and vehicle-to-infrastructure connectivity, automation, and others, poses a challenge of not only understanding the effects of each technology individually, but also how each interacts in future scenarios when these technologies co-evolve [2]. Generally, such technologies can provide benefits such as lower vehicle ownership [3], reduced congestion due to higher traffic capacity [4], and more efficient travel through connectivity and intelligent transportation system management [5], [6]. Nevertheless, such technologies can induce demand and lead to adverse outcomes through higher travel with low occupancy or unoccupied vehicles [7], lower overall capacity due to excessive pickup and drop-offs at curbs [8], and higher travel due to increased accessibility [9].

In this study, we leverage an agent-based activity-based simulation model called POLARIS [10] that has been integrated into a regional modeling workflow [11] to jointly explore multiple policy aspects, and apply them to the Chicago metro area. The paper is organized as follows: the next section provides the methodology detailing the agent-based tool and assumptions on the policy levers; followed by a summary of the data model, and finally concluding with results and a discussion.

## METHODOLOGY

This study employs the POLARIS agent-based-modeling workflow [10], [11] to analyze the impacts of the selected technologies on the Chicago region. The main components of the workflow can be seen in Figure 1. Previous deployments of the POLARIS model have looked at individual cities [12], or individual technologies [13], [14], [15].

POLARIS is a full-featured agent-based model that includes a population synthesizer and a comprehensive activity-based model on the demand-side and traffic and transit simulators along with dynamic route assignment, TNC operation and electric vehicles charging in the supply-side. Therefore, POLARIS can capture demand and supply aspects in an integrated manner [10], [14].

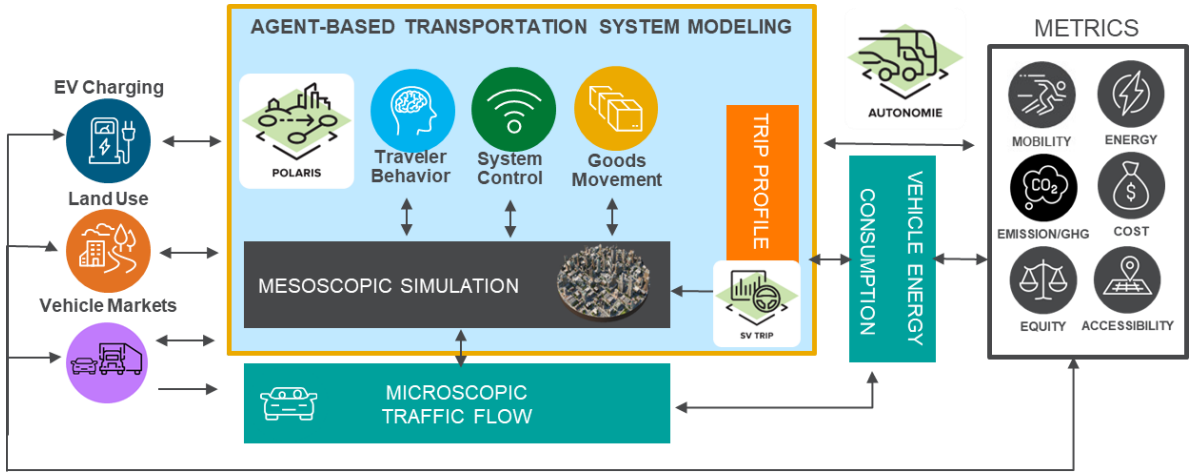

*Figure 1. SMART Mobility Modeling Workflow*

POLARIS implements an activity-based modeling approach using the ADAPTS framework [16]. Travelers make decisions at different timescales - long-term, mid-term and within-day - as needed. Work/School location and household vehicle choices are the two key long-term choices within POLARIS. These long-term decisions constrain the short-term decision-making process. The short-term decisions include the set of activities to complete and the planning of each activity. Every activity has an associated type (work, personal, errands, etc.), meta-characteristics such as flexibility and planning horizon, and individual attributes such as location, start time, duration, and the preferred travel mode. The activity scheduler constructs the daily plan by making use of sub-modules for mode, destination, and departure-time choice that are calibrated for local conditions. The schedule is then carried out over the course of the day and subject to changes in the face of unexpected changes in travel times or vehicle unavailability. The replanning process includes conflicting resolution for cases where multiple agents are competing for the same resource [17].

POLARIS also features a mesoscopic implementation [18] of the link-transmission-model [19]) for traffic flow simulation. The model has adaptations to the LWR model to track vehicles individually. The key modification is to the node models to ensure discrete flows at every time-step [21]. In this work, we change the link model into Lagrangian Coordinates so that the position and speed of every vehicle is known at every time step [22]. It also allows the implementation of adaptive signal control algorithms that account for the state of each vehicle within the link. The flow computation follows [18], which contains further detail on the process.

A computationally efficient multi-modal router was implemented to simulate the millions of trips expected for any moderately sized region, since each trip requires a router request to compute the best path for the agent at the specific departure time, along with any re-routes necessary due to congestion formation or changes in the activity schedule. POLARIS features a point-to-point time-dependent A* multimodal router [23], that considers a generalized cost structure with turn travel times as its major component. The expected generalized cost on a turn is the weighted average between prevailing and expected travel times at the expected arrive time at that link. The expected future travel times are informed by past iterations. An information-mixing dynamic traffic assignment adapts the weights on prevailing condition and expected travel times based on the iteration number and the distance from that link. For further information on the dynamic traffic assignment approach, we refer to [23]

Transit operations are also simulated with high fidelity in POLARIS [24]. Each individual transit trip is simulated and agents can board, alight, and transfer across routes to reach their destination. The access and egress walking legs are also considered. Agents can also perform inter-modal trips such as walk-to-transit, park-and-ride. The movement is informed by the time-dependent multimodal router [23].

POLARIS additionally simulates the operation of Transportation Network Companies (TNC) and freight delivery firms. Requests for shared-rides from TNCs are an outcome of the mode-choice model and the TNC operators undertake actions to pickup and dropoff passengers according to the current vehicle size, possibility of pooling, operational strategy and other aspects. For further information on TNC operations, we refer to [25] . The freight activity model in POLARIS is implemented as an agent-based model from the perspective of freight involved firms, including shippers, receivers and logistics companies. Connections are made between firms and logistics channels are selected based on the CRISTAL framework, to represent business-to-business shipments in a region. Business-to-consumer movements are simulated in a parcel delivery module to satisfy e-commerce with optimized routing, on-demand delivery for grocery and prepared mail requested by household agents in POLARIS. For details on freight simulation see [26], [27], [28]).

Finally, to estimate vehicle energy consumption, Autonomie [29], a vehicle system simulation tool utilized in automotive control-system design, simulation, and analysis is connected to POLARIS. This tool assesses the energy consumption of various advanced powertrain technologies. Autonomie includes vehicle models to evaluate energy consumption during simulated trips. These models include conventional powertrains such as gasoline or diesel internal combustion engines (ICE), belt-integrated starter-generator (ISG), hybrid electric vehicles (HEVs), plug-in hybrid vehicles (PHEVs), battery electric vehicles (BEVs), and fuel-cell vehicles (FCVs) across various light-duty, medium-duty, and heavy-duty vehicle classes [30]. Furthermore, Autonomie allows the integration of connectivity for these vehicles and accommodates different automation levels from L0 to L4. In addition to consumption estimation, Autonomie assesses emissions for each vehicle during the trips. Emission rates from GREET are integrated into Autonomie to estimate emissions. Subsequently, GHG emissions are computed using GREET well-to-wheel emission factors and Autonomie's vehicle consumption estimation.

The POLARIS model has been extensively calibrated and validated for multiple model regions [12], [31], [32] and applied to multiple mobility technology applications [11], [12], [33]

## SCENARIO DESIGN

The study was conducted for the Chicago region using the workflow described above after extensive consultation with local stakeholders to gauge policies of interest that would likely be of interest to explore in the mid-term, with a forecast horizon of approximately 20 years. The results of these discussions, and the demand and supply settings that were determined for this analysis are shown in Table 1**Error! Reference source not found.**, and described in detail in the following section. It was decided by the study team to implement a design where two levels for each scenario lever were set, generally either an off/on or low/high combination, except for vehicle electrification where three levels were selected. Table I denotes the different scenarios studied by using a variety of levers related to demand and supply factors. A total of 192 scenarios were tested with six demand-related scenarios combined with 32 supply-related scenarios, in a full-factorial design over all the levers. The following subsections define the implementation of these policies.

*Table 1. Study design*

| | Supply | | Demand |
|---|---|---|---|
| **Congestion pricing** | Fixed time-of-day tolls based on expected delay for all interstates. | **Telecommuting** | Fixed at 15% daily telecommuting. |
| **Transit Investments** | Increase speed on 39 CTA routes, increase frequency 40% on Pace/METRA. | **Vehicle Connectivity / Automation** | Fixed at 40% penetration of L2-CACC. |
| **Connected Infrastructure** | Connected signals at all Major intersections that interact with CV (60% of total signals). | **E-Commerce / On-demand delivery** | Increase over time 2% y-o-y versus 6%. |
| **Rideshare / First-mile last-mile (FMLM)** | Subsidize FMLM to transit stops outside Chicago and impose C2C/parking policies on regular trips. | **Vehicle electrification** | Light duty (LD) vehicles: 15%, 35%, 75% Heavy duty (HD) vehicles: 10%, 25%, 50% |
| **Freight** | Off-hours delivery (5% vs. 15%) acceptance. | | |

## Supply Levers

POLARIS allows both dynamic pricing and a fixed time-dependent price for each link with any time interval. A fixed time-dependent pricing on expressways was selected for study, as this option is more feasible for near-term deployment from a technical and public support standpoint. Aiming at reducing peak congestion, fixed prices were set based on average delay per distance observed for each time step. The conversion from delay into monetary cost is performed by assuming $18/hour of value of time. Since the delay and not the travel time is priced, there is therefore no price incurred if no delay is experienced in the link. The tolls average $0.16 per km in the AM peak, $0.14 per km in the PM peak, with substantially lower values in mid-day and off-peak periods. The response to different prices is not limited to different route choice. A change in the price can lead to different destination, mode and departure time choices. Although it is expected a lower VMT at peak on expressways, the net-impact including the arterial and local roads are unclear due to its multiple travel behavior impacts. Pricing expressways may shift traffic to arterial and local roads that have lower capacity and free flow speed which in turn would lead to higher VHT and lower average speed. On the other hand, the higher price on the peaks may shift the demand away from the peak thereby reducing overall delay. Aside from the congestion impacts, congestion pricing also interacts with transit. In the scenarios with transit investments enabled, 20% of the revenue from congestion pricing is added to the transit investment budget, while the rest is assumed to be rebated to low-income residents.

The Chicago region has three major transit agencies: Chicago Transit Authority (CTA) for urban bus and urban rail, Pace for suburban bus, and Metra for commuter rail services. Based on several studies and discussions with the CTA and the Regional Transportation Authority (RTA), the following service improvements were implemented: 30% speed increase on 39 CTA bus routes which also results in ~30% frequency increase, 40% across-the-board frequency increase for Pace, and ensure a 30- minute headway on every Metra route between 6 am and 10 pm which results in 40% overall frequency increase. Moreover, Bus Rapid Transit (BRT) routes are added on Ashland and Western avenues for CTA, and express (limited stop) bus routes on the 95th, Dempster, and Halsted streets for Pace. Finally, CTA's highest-ridership rail route, the Red Line, is extended by four stations on the Southern end of the route, from 95th Street to 130th Street.

The connected infrastructure policy is implemented through a local signal control strategy that leverages the knowledge of speed and position of the connected vehicles in the approaching links that operates similar as an actuated control. Each phase has its minimum, desired, and maximum green times. After the minimum green time has elapsed, an early switch to the next phase occurs if no vehicle on the protected movements can cross in the next $T_g$ seconds and there exists at least one vehicle from opposing movements that can cross the intersections in time $T_r$. The desired green time is performed if there are in both protected movements of the prevailing phase and also in any of the other movements or in the opposite case in which the intersections sense no vehicles in any movement at all. Finally, the prevailing phase may hold if its protected phases have incoming vehicles and there are no incoming vehicles at the pre-defined periods in the other movements. For this study we set $T_g$ =6s and $T_r$ =3s.

A fleet of shared vehicles is also included to reflect the ride-hailing service in the area. For the future scenario, two sets of operators are defined to accommodate the variable nature of trip density in the area, and consistent with local transit-agency feedback. In the downtown area, in and around the central business district (CBD), a corner-to-corner routing algorithm is required to minimize the fleet travel distance that would otherwise result from pickups in a door-to-door service. The matching algorithm implemented by [34] for virtual stops and walking legs can perform especially well in the gridded network typically found near downtowns. In areas with less dense trips, housing, and road network, a first-mile-last-mile focused approach is taken. The fleet of operators serving outside the CBD area are all assumed available to support first-mile-last-mile (FMLM) trips to and from a transit stop. A router-based shortest path is used to identify FMLM feasibility efficiently as documented by [35]. These trips, are also provided with a subsidy of 50% of the typical ride-hailing fare as a scenario lever to encourage users to hail a ride to a commuter or other transit stop in the suburb to access the typically congested downtown areas.

The final supply management strategy is to incentivize off-hours delivery (OHD), as it is potentially beneficial in reducing delivery costs as trucks are usually traveling off-peak times or at night lowering congestion and fuel consumption. In the study, business-to-business OHD was implemented based on a behavioral model from the literature [36]. Since OHD is a current practice for the shippers, 5% of the companies are assumed as already receiving OHD as a baseline. For the scenario alternative, OHD acceptance was increased to 15% among receivers. The deliveries were shifted from daytime to overnight (7 PM-6 AM). Trips for medium and heavy-duty vehicles increased about 75% overnight in the scenario, compared to baseline.

**Demand Levers**

Various potential demand levers were also explored in the study. In order to limit the design of experiments scale, two of the levers, telecommuting and CACC, were set at fixed, forecast values as these have been extensively studied in previous work [12], [37]. First, the telecommuting level was fixed at 15% of workers telecommuting on the study day. To understand the effects of the telecommuting, a behavioral model was estimated [38] and implemented in the POLARIS to generate telecommuting choices and simulate the changes in workers' daily activity-travel pattern. The model first determines whether a worker telecommutes, and if so heuristics were applied to replace the work activity with an at-home work activity and adapt the worker's out-of-home activities. CACC vehicle adoption impact was also set at a fixed 40% penetration rate for this study. CACC leads to higher roadway capacity and shorter time gaps on freeways. This is captured by adjusting the shock-wave speed and link capacity for each link depending on the share of CACC-enabled vehicles currently on that link. This framework has been applied before [12], [39], and it assumes a quadratic increase in capacity following the results in [40]. A technology choice model that models the likelihood that an individual adopts AVs based on a number of explanatory variables such as income and household composition was used [42] to distribute the CACC vehicles to individual households.

E-commerce refers to the buying and selling of goods and services over the internet and delivered by the parcel delivery companies (e.g., UPS, FedEx, USPS, and Amazon). E-commerce sales have grown from 1% of US retail purchasing in 2000 to about 12% in 2020 (pre-pandemic era), then sales jump to 16% during the pandemic and now stabilized to 15% by 2023 [43]. Similarly, COVID-19 social distancing policies and the growing demand for online purchases of goods accelerated the move towards on-demand delivery (ODD) of prepared meals and groceries. The shared mobility model in POLARIS [14] is updated to serve ODD requests during the simulation. ODD requests are passed on to the operator by the household agent at the time assigned during the simulation. Heuristic dispatching algorithms are used to match vehicles to requests and complete the delivery. To take the growing online demand into account, we defined two scenarios where we analyze a low or conservative grow of 2% year to year (y-o-y) versus a high demand scenario with 6% y-o-y. With these rates, the approximate demand for the low case corresponds to 3.5 (e-commerce), 0.7 (ODD groceries), and 1.8 (ODD meals) deliveries per household per week. While the high scenarios correspond to 9.8 (e-commerce), 1.5 (ODD groceries), and 3.5 (ODD meals) deliveries per household per week.

Finally, the last demand-related setting was the vehicle electrification level in the region. As a critical element of the national transportation decarbonization strategy which is being intensively deployed and incentivized it is important to understand how transportation electrification interacts with other potential future supply interventions and under various demand related scenarios. As such, three setting levels were selected for analysis, representing a low, moderate and high level of EV penetration in the passenger and freight markets by 2040. The light duty (LD) vehicle levels were set at 15%, 35%, and 75% penetration in vehicles stock respectively, while the associated heavy duty (HD) vehicles were set at 10%, 25%, and 50%, representing the increased difficultly of electrifying this segment. These marginal totals were then distributed across existing detailed vehicle registration distributions by class, segment, powertrain and vintage at the zip code level, to generate the forecast distribution of household and freight vehicles for the model runs.

**Chicago Model**

The scenarios defined above were executed in POLARIS on Chicago area in the United States. It corresponds to 48.4k links, 35.9k nodes,18.6k total center line miles, and comprises a population of 10.4M and 40.8M trips per day. The city core has one of the most comprehensive transit infrastructures in the US. Suburban Chicago is mostly car-dependent, with some access provided by suburban rail. This model has been developed and has been actively being maintained to allow cross comparison with other cities and disentangle the impacts of new mobility technologies from aspects such as socio-economic characteristics, land-use, and existing transit infrastructure. The calibration of the baseline case was carried out using travel surveys collected by the MPO. The MPOs also provided the trip OD tables for external-external, external-internal, and internal-external trips which we synthesized into individual trips.

## RESULTS AND DISCUSSION

The following section describes the main analysis findings from running the large-scale simulation design of experiments. It should be emphasized here that with a total of 192 different scenarios that were run over 8 times each at converged settings, there are over 1,600 individual region-level sets of outcomes, each reporting on 20+ core metrics. To enhance interpretability and present the findings in a meaningful way, multivariate analysis was performed to generate a regression model for each key metric at the region level to inform on the statistical relationships between scenario settings and key performance outcomes, and to identify optimal combinations of settings depending on variable goals and priorities.

### Mobility Analysis

A multivariate analysis is used to explore the impact and interaction of the different scenario levers studies here. Table 2 shows an analysis of the system VHT for ground traffic. Results indicate that demand levers such as OHD and e-commerce/ODD policies help in reducing VHT, with e-commerce/ODD having the highest impact with 4.6% reduction due to replacing individual shopping trips with optimized deliveries. Similarly, congestion pricing and signal coordination can also help reduce the system VHT by 0.9% and 1.7% respectively.

*Table 2. Regression analysis of system VHT (million hours) with levers*

| Variable | Sensitivity | Coef. | Std Err | t | P>|t| | Signif. |
|---|---|---|---|---|---|---|
| Constant | | 6.975 | 0.035 | 198.665 | 0.000 | *** |
| Congestion Pricing | -0.9% | -0.061 | 0.036 | -1.703 | 0.089 | . |
| Signals | -1.7% | -0.119 | 0.040 | -3.016 | 0.003 | ** |
| TNC | 1.0% | 0.071 | 0.039 | 1.790 | 0.074 | . |
| OHD | -0.4% | -0.029 | 0.016 | -1.791 | 0.074 | . |
| E-commerce/ODD | -4.6% | -0.321 | 0.032 | -9.957 | 0.000 | *** |
| Pricing X Signals | 1.9% | 0.132 | 0.046 | 2.889 | 0.004 | ** |
| Pricing X Signals X TNC | -1.5% | -0.108 | 0.065 | -1.668 | 0.096 | . |
| E-commerce\ODD X Transit | 1.6% | 0.110 | 0.032 | 3.393 | 0.001 | *** |
| E-commerce\ODD X TNC | -0.8% | -0.053 | 0.032 | -1.653 | 0.099 | . |

Signif. Codes: 0 '***' 0.001 '**' 0.01 '*' 0.05 '.' 0.1 ' ' 1          N = 760, Adj. R2 = 0.364

Importantly, many of the interaction terms between the scenario levers exhibit statistical significance, demonstrating the importance of this type of scenario experiment. Substantial interaction happens between the pricing, signal and TNC policy levers, the e-commerce and transit lever, and the e-commerce/on-demand delivery and TNC policy levers. This can lead to important findings that would not be observed while exploring each policy in isolation. For example, while TNC policies can lead to a 1% increase in VHT, largely due to the subsidized FMLM fares increasing TNC usage more than the efficient corner-to-corner policy decreases it, it has beneficial interactions with the pricing, signals and e-commerce/ODD levers. The ODD interaction is straightforward as more vehicles in the system that are handling ODD deliveries allow for more efficiency when handling rideshare requests and vice versa. The interaction between signals, TNC and pricing is more complicated, but largely is due to congestion pricing and FMLM subsidy synergy to get more travelers in the TNC and ride mode. These results can be seen in Figure 2, which illustrates the effect of the different levers on VHT and shows the optimal combination of policies for the maximum reduction of VHT. With the optimal scenarios of policies, VHT can be reduced up to 7%. So, in spite of TNC in isolation having a detrimental impact on VHT it is needed to reach the overall optimal reduction of 7%. A similar effect happens in the other direction for pricing and signals, with the levers in isolation reducing VHT by 0.9 and 1.7%, but when combined they have a worse outcome than each in isolation with a 0.7%. Again, though, both are necessary to reach the optimal outcome.

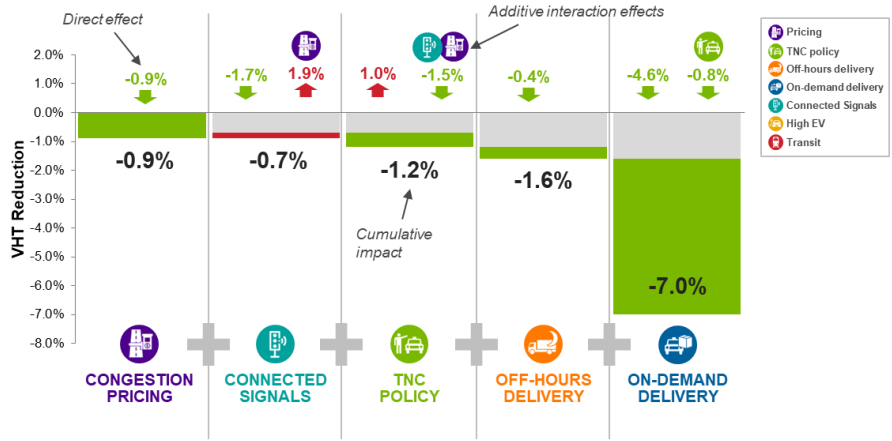

*Figure 2. Individual and Cumulative Impact on System VHT from Optimal Interventions*

**Energy Analysis**

The energy analysis is summarized with the impact of levers on the energy use (in GWh/day units) as shown in Table 3. Regression analysis of energy use and GHG interaction with levers. Results suggest that electric vehicle deployment is critical for energy reduction. A medium EV penetration rate could lead to a 27% reduction on energy use while a high EV penetration rate could lead to 35% reduction, compared to a low rate. Congestion pricing policies, OHD and rideshare policies could also help reduce energy consumption. While a combination of rideshare and e-commerce could lead to higher energy consumption. Policies focusing on increasing EV penetration could significantly improve long-term energy consumption.

An analysis of the normalized GHG emission (g/mi/day) is shown in Table 3. Vehicle electrification could lead to a 30% (medium EV) and 40% (high EV) reduction on GHG emissions. Similarly, congestion pricing, signal coordination and OHD policies have a positive effect on decreasing emissions. Congestion pricing, signal coordination, and OHD policies could be used for transport decarbonization.

*Table 3. Regression analysis of energy use and GHG interaction with levers*

| | Variable | Sensitivity | Coef. | Std Err | t | P>\|t\| | Signif. |
|---|---|---|---|---|---|---|---|
| Energy Use (GWhr) | Constant | | 331.848 | 0.697 | 476.290 | 0.000 | *** |
| | Congestion Pricing | -1.50% | -4.977 | 0.441 | -11.294 | 0.000 | *** |
| | FMLM/TNC | 0.43% | 1.412 | 0.623 | 2.265 | 0.025 | * |
| | E-commerce/ODD | 0.30% | 1.007 | 0.441 | 2.286 | 0.023 | * |
| | EV (Medium) | -27.25% | -90.428 | 0.540 | -167.556 | 0.000 | *** |
| | EV (High) | -35.33% | -117.236 | 0.540 | -217.228 | 0.000 | *** |
| | OHD X FMLM/TNC | -0.75% | -2.474 | 0.881 | -2.807 | 0.006 | ** |

| | | Sensitivity | Coef. | Std Err | t | P>|t| | Signif. |
|---|---|---|---|---|---|---|---|
| **GHG (g/mi)** | Constant | | 449.591 | 0.630 | 713.633 | 0.000 | *** |
| | Congestion Pricing | -0.73% | -3.301 | 0.420 | -7.859 | 0.000 | *** |
| | Signal Coordination | -0.32% | -1.461 | 0.420 | -3.477 | 0.001 | *** |
| | OHD | -0.25% | -1.116 | 0.420 | -2.657 | 0.009 | ** |
| | E-commerce/ODD | 0.84% | 3.786 | 0.420 | 9.015 | 0.000 | *** |
| | EV (Medium) | -28.66% | -128.854 | 0.514 | -250.497 | 0.000 | *** |
| | EV (High) | -39.95% | -179.629 | 0.514 | -349.204 | 0.000 | *** |

## System Efficiency

The changes in travel efficiency driven by various combinations of scenario levers can also serve as a useful metric and can be calculated in multiple ways. A general efficiency metric determined by the change in the amount of energy needed to travel one productive mile (i.e. travel to reach an activity, move a passenger or deliver goods) and a productivity metric, Mobility Energy Productivity [44], which quantifies accessibility to activity opportunities were calculated here. When looking at system efficiency in productive miles per kilowatt hour of energy expended, many of the individual levers are effective at increasing efficiency and there are positive synergistic effects that improve efficiency beyond just electrifying vehicles. In the optimal combination of high electrification, congestion pricing, transit and off-hours delivery, the EV lever contributes most of the improvement at 46%. However, pricing, transit and OHD combined contribute another 7% to the optimal 53% efficiency improvement, demonstrating the importance of transportation systems and mobility to achieving decarbonization goals.

*Table 4. Regression analysis of system efficiency (mi/kWhr) and interaction with levers*

| Variable | Sensitivity | Coef. | Std Err | t | P>|t| | Signif. |
|---|---|---|---|---|---|---|
| Constant | 1.1% | 0.874976 | 0.002533 | 345.3734 | 0.00 | *** |
| Congestion Pricing | -3.2% | 0.010787 | 0.003119 | 3.458468 | 0.00 | *** |
| Ecomm/ODD | 32.3% | -0.0317 | 0.002751 | -11.5243 | 0.00 | *** |
| EV_MED | 46.0% | 0.323262 | 0.003119 | 103.64 | 0.00 | *** |
| EV_HIGH | 0.8% | 0.460242 | 0.004342 | 106.0081 | 0.00 | *** |
| Congestion Pricing_Off-Hours Freight | 1.7% | 0.008452 | 0.003602 | 2.346768 | 0.02 | * |
| Congestion Pricing_EV_HIGH | 1.1% | 0.017438 | 0.004411 | 3.953343 | 0.00 | *** |
| Transit_EV_MED | 1.8% | 0.011275 | 0.003602 | 3.130576 | 0.00 | *** |
| Transit_EV_HIGH | 0.5% | 0.0182 | 0.003602 | 5.053369 | 0.00 | *** |
| Rideshare/FMLM_Off-Hours Freight | -0.7% | 0.005282 | 0.002751 | 1.920123 | 0.06 | . |
| Off-Hours Freight_Ecomm/ODD | 0.9% | -0.00658 | 0.003602 | -1.82821 | 0.07 | . |
| Off-Hours Freight_EV_HIGH | 1.1% | 0.009159 | 0.004126 | 2.219693 | 0.03 | * |

## Equity and Environmental Justice

Equity and environmental justice impact of the transportation solutions are a critical element of current decarbonization planning efforts, to ensure that any plan does not have disproportionate impacts on disadvantaged groups and that any benefits or burdens are distributed equitable. Results of the simulated travel cost per capita in dollars per day were estimated by income quantiles. Results show that the average travel cost per capita is $14 (bottom), $19 ($2^{nd}$), $26 ($3^{rd}$), $32 ($4^{th}$), and $38 (top), where the cost includes vehicle operating costs, any fares tolls or parking charges, and an estimate of the time cost based on the individual travel value of time. The percentage change from the BAU scenario was estimated for all levers to assess the cost burden changes across the different income quantiles. The results show that the greater disparity in changes corresponds to

scenarios with congestion pricing policies and those where FMLM is not subsidized. A multivariate analysis of the impact of levers on the bottom and top quantile is shown in Table 5. The results highlight that congestion pricing has a greater sensitivity and tends to affect more to the population from the bottom income percentile, while the high-income percentile benefits the most from signal coordination. A combination of both congestion pricing and signal coordination policies could increase the cost burden for both population groups. Share mobility and subsidized FMLM could help reduce cost burden by 7% for the bottom income population, while its effect is not significant for the top income population as expected due to the much heavier reliance of low-income households on shared travel modes and the higher share of overall income the that fixed fares for those modes represent. Similarly, E-commerce and ODD could help reduce the cost burden for all groups by 7% (bottom) and 5% (top) through reducing the need to travel.

*Table 5. Regression analysis of cost burden by lever for bottom and top income quantiles*

| Variable | Bottom Percentile | | Top Percentile | |
|---|---|---|---|---|
| | Sensitivity | P>|t| | Sensitivity | P>|t| |
| Congestion Pricing | 1.00% | 0.011 | 0.82% | 0.166 |
| Signal Coordination | -0.65% | 0.098 | -2.40% | 0.000 |
| FMLM/TNC | -6.71% | 0.000 | -0.25% | 0.547 |
| Transit | 1.43% | 0.000 | 0.96% | 0.022 |
| E-commerce/ODD | -7.02% | 0.000 | -4.70% | 0.000 |
| Pricing X Signals | 0.97% | 0.081 | 1.71% | 0.042 |
| E-commerce/ODD X OHD | -0.71% | 0.072 | -0.89% | 0.135 |

It is also useful to explore the impact on environmental justice spatial throughout the region, although necessarily limiting it to one scenario at a time. Exploring EEJ spatially is a fundamental component of the Justice40 initiative [45]. The J40 initiative includes the spatial analysis and classification of Disadvantaged Communities (DAC) against which we can explore the differential impact of the scenarios studies here. Generally, DACs are defined as Census Tracts which have greater that the 65th percentile of low-income households and also a 90th percentile or greater outcome on a number of key metrics including health, pollution, energy, transportation and so on. When we plot the reduction in PM2.5 exposure by traffic analysis zone observed in the optimal scenario in the Chicago model against the DACs in the region we see results as shown in Figure 1. The figure shows the differential impact that the optimal emissions reduction scenario identified from the results in Table 3 would give. In this case, the optimal scenario combines congestion pricing, signal coordination, off-hours delivery, and high electrification. In that combination, it was found that DACs on average have a reduction of approximately 0.3g/person of PM2.5 exposure, while non-DAC have a 0.14 g/person reduction, with the scenario overall reducing PM2.5 emissions by 24%. So, in this case the combination of policies seems to be effective at equitably improving this key metric. In the graph, much of the PM2.5 reduction occurs along either major highways, or substantial freight distribution or other heavily industrialized areas, which tend to both be in disadvantaged communities, although substantial improvements are seen in non-DAC as well.

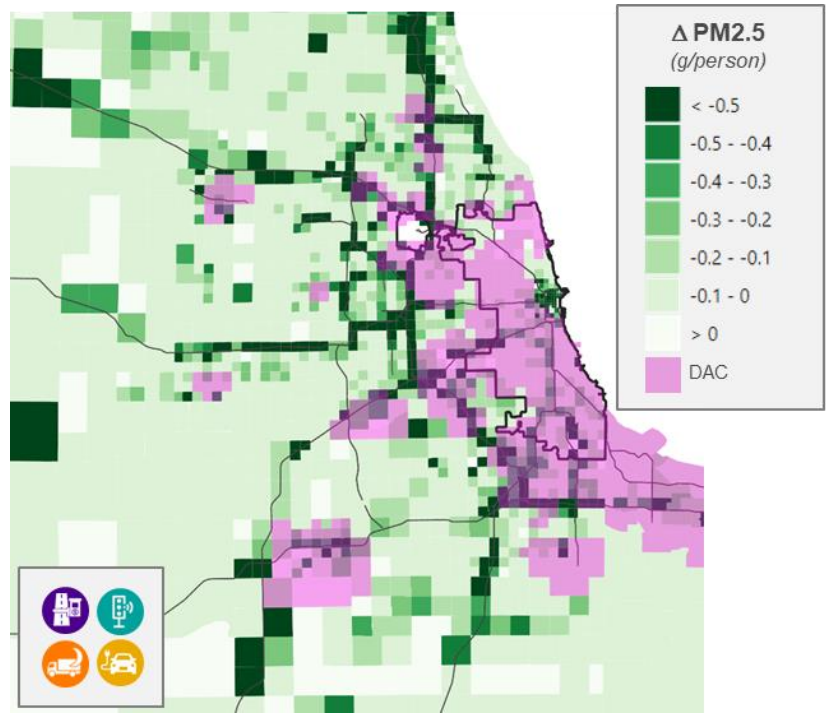

*Figure 3. Reduction in PM2.5 exposure by TAZ*

## Selecting Optimal Scenario Combinations

The preceding discussions have all focused on identifying the impacts on individual metrics from the lever combinations through regression analysis and using the results to identify the optimal combination of that metric. However, in many instances, there will not be one target metric to optimize, but rather a series of competing priorities that could inform project selection. To evaluate the impact on key outcomes from prioritizing different sets of metrics, an integer program was developed using selection of different levers (setting to 0 or 1) as the decision variables and the target function determined as a weighted combination of all of the expected metric outcomes determined using the regression results for a total of six different metrics (VMT, VHT, energy, GHG, efficiency and MEP), with the weights being configurable to represent different priorities. The results for two possible prioritizations are shown: the first is an energy focused setting, where energy, GHG, efficiency and MEP change are normalized and given equal weight, and the second is a mobility focused scenario where the same process Is applied to the VMT, VHT and MEP metrics. We can see that the optimal lever settings are slightly different in each case, where the energy-focused result includes transit, while the mobility results trades that for off-hours delivery. This is likely due to the fact that OHD is generally more beneficial in mobility metrics as it replaces a substantial amount of travel miles due to more efficient delivery tours, but this benefit is reduced when looking at energy as these tours are implemented with less-efficient medium duty freight vehicles – a result which is further emphasized in the high-EV scenarios. Both solutions give fairly good results across most metrics, but the largest difference is seen in the VHT measure, which is likely to be a key performance metric for transportation authorities. This demonstrates the complexity of this decision-making process which analyses and platforms like POLARIS can help to support.

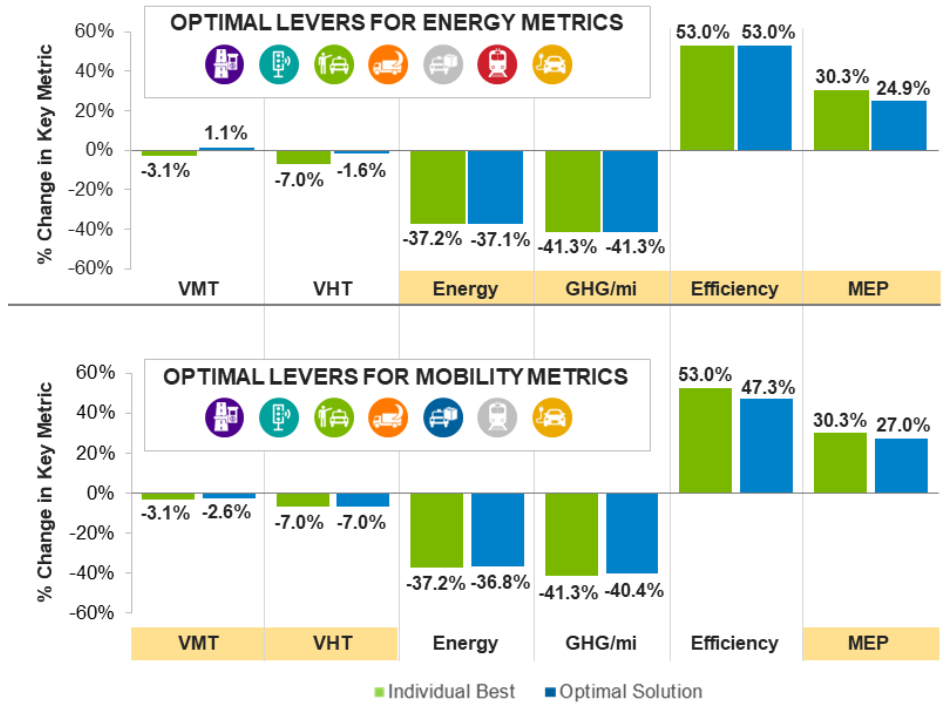

*Figure 4. Optimal Levers and Changes in Key Metrics for Different Priority Inputs*

## SUMMARY AND CONCLUSIONS

In this study, scenarios were simulated for the Chicago area to identify impacts on congestion, travel distance, energy use, emissions, equity, and overall efficiency. Different supply and demand levers were defined and applied in a design of experiments that involved engagement with local stakeholders. Large scale simulation design and analysis is relevant for the evaluation of policies at scale, especially to address challenging systemic issues like transportation decarbonization. The principal finding of this work suggest that, while independent isolated policies can have significant impact on improving the system efficiency and reducing energy impacts, our multivariate statistical analysis and optimal solution assessment highlight the need for understanding and accounting for policy interactions through multivariate impact assessment. The optimal combination of scenarios was found to lead to up to a 7% reduction in system VHT. Vehicle electrification is critical for reaching decarbonization goals, and high EV penetration rates can lead to a reduction of 35% in energy use and 40% in GHG emissions. In terms of system efficiency in productive miles per kilowatt hour of energy expended, the EV lever and pricing, transit and OHD combined contribute to a 53% improvement, with 14% of that improvement being driven by policies and technology deployments unrelated to electrification. In terms of EEJ, a spatial assessment of impact of the optimal solution found that the combination of policies seems to be effective at equitably reducing PM2.5. This work is critical to support the objective of transportation decarbonization, while maintaining system performance and accounting for transportation equity and EEJ.

Some limitations of the current study surround the lack of detailed household vehicle fleet evolution, which necessarily limits the response of some policies intended to drive modal shifts, namely FMLM subsidies and transit enhancement. In those cases, the lack of response of households to the changing conditions due to the policy interventions in terms of vehicle disposal, means that households which own personal vehicles generally continue to use them. This limitation has motivated work on vehicle transaction modeling and vehicle holdings, especially in response to future mobility technologies [46], [47], which can be incorporated into future studies. Additionally, many of the scenario levers in this study would either benefit from land use policies that support their goals, e.g. transit-oriented development land use policy to enhance the transit improvement lever, or would be expected to potentially drive land use changes, e.g. congestion pricing, telecommuting, etc. Since land use growth related to the specify policies was not accounted for here, there is potential for outcomes to be better or worse than expected here, depending on the interaction with the land use system. The POLARIS modeling workflow does include linkage to land use modeling through UrbanSim [48], although adding the temporal dimension of land use and transportation change over time would add significantly to the computational burden of the study. For example, even with a five year update period, to run the scenarios here out to 2040 from the baseline year of 2020 would increase the number of simulation runs by five times. Therefore, the selected pathway was to use this study to explore a large set of scenario combinations and explore in more detail the most promising pathways with full land use integration over time.

## ACKNOWLEDGEMENTS

The work done in this paper was sponsored by the U.S. Department of Energy (DOE) Vehicle Technologies Office (VTO) under the Systems and Modeling for Accelerated Research in Transportation (SMART) Mobility Laboratory Consortium, an initiative of the Energy Efficient Mobility Systems (EEMS) Program. Erin Boyd, DOE technology manager, has played important roles in establishing the project concept, advancing implementation, and providing ongoing guidance. We also acknowledge our stakeholder contributions in terms of guidance, feedback and data from CTA, CDOT, RTA and CMAP. Also, we gratefully acknowledge the computing resources provided on Bebop, a high-performance computing cluster operated by the Laboratory Computing Resource Center at Argonne National Laboratory. Finally, we thank all our university collaborators who have contributed to the program throughout the SMART project.